\documentclass[pra,twocolumn,showpacs]{revtex4-2}

\usepackage{amsmath}
\usepackage{latexsym}
\usepackage{amssymb}
\usepackage{graphicx}
\usepackage[colorlinks=true, citecolor=blue, urlcolor=blue]{hyperref}
\usepackage{float}
\usepackage{amsfonts}
\usepackage{textcomp}
\usepackage{mathpazo}
\usepackage{comment}
\usepackage{xr}

\usepackage{bbm}

\usepackage{xcolor}
\definecolor{myurlcolor}{rgb}{0,.5,.5}
\definecolor{mycitecolor}{rgb}{0,.6,0}
\definecolor{myrefcolor}{rgb}{2,0,0}
\usepackage{hyperref}
\hypersetup{colorlinks,
linkcolor=myrefcolor,
citecolor=mycitecolor,
urlcolor=myurlcolor}

\usepackage[draft]{fixme}
\usepackage{amsmath,bbm}
\usepackage{graphicx}
\usepackage{amsfonts}
\usepackage{amssymb}
\usepackage{amsmath, amssymb, amsthm,verbatim,graphicx,bbm}
\usepackage{mathrsfs}
\usepackage{color,xcolor,longtable}

\usepackage{xr}
\makeatletter
\newcommand*{\addFileDependency}[1]{
  \typeout{(#1)}
  \@addtofilelist{#1}
  \IfFileExists{#1}{}{\typeout{No file #1.}}
}
\makeatother

\newcommand*{\myexternaldocument}[1]{
    \externaldocument{#1}
    \addFileDependency{#1.tex}
    \addFileDependency{#1.aux}
}

\myexternaldocument{manuscript_PRL}

\newcommand{\beq}[0]{\begin{equation}}
\newcommand{\eeq}[0]{\end{equation}}

\newcommand{\one}{\leavevmode\hbox{\small1\normalsize\kern-.33em1}}

\def\be{\begin{equation}}
\def\ee{\end{equation}}
\def\ben{\begin{eqnarray}}
\def\een{\end{eqnarray}}
\def\eea{\end{array}}
\def\bea{\begin{array}}

\newcommand{\Tr}[1]{\mathrm{Tr}#1}
\newcommand{\bei}{\begin{itemize}}
\newcommand{\eei}{\end{itemize}}
\newcommand{\ket}[1]{|#1\rangle}
\newcommand{\bra}[1]{\langle#1|}

\newcommand{\proj}[1]{\ket{#1}\!\!\bra{#1}}

\newcommand{\I}{\mathbbm{1}}

\renewcommand{\emph}[1]{\textbf{#1}}

\makeatletter
\newtheorem*{rep@theorem}{\rep@title}
\newcommand{\newreptheorem}[2]{%
\newenvironment{rep#1}[1]{%
 \def\rep@title{#2 \ref{##1}}%
 \begin{rep@theorem}}%
 {\end{rep@theorem}}}
\makeatother

\theoremstyle{plain}
\newtheorem{thm}{Theorem}
\newtheorem*{thm*}{Theorem}
\newreptheorem{thm}{Theorem}

\newtheorem{defn}[thm]{Definition}

\theoremstyle{definition}

\theoremstyle{remark}

\usepackage[T1]{fontenc}

\begin{document}

\title{Shared randomness allows violation of macroscopic realism using a single measurement}
\author{Shubhayan Sarkar}
\email{shubhayan.sarkar@ulb.be}
\affiliation{Laboratoire d’Information Quantique, Université libre de Bruxelles (ULB), Av. F. D. Roosevelt 50, 1050 Bruxelles, Belgium}

\begin{abstract}	
Macro-realistic description of systems is based majorly on two basic intuitions about the classical world, namely, macrorealism per se, that is, the system is always in a distinct state, and non-invasive measurements, that is, measurements do not disturb the system. Given the assumption of no-signalling in time, one utilizes Leggett-Garg inequalities to observe a violation of macroscopic realism which requires at least three measurements. In this work, we show that if one has access to shared randomness then one can observe a violation of macroscopic realism using a single measurement even if no signalling in time is satisfied. Interestingly, using the proposed scheme one can also rule out a larger class of models, which we term {\it{macroscopic no-signalling}} theories which can not violate the no-signalling in time conditions. We further construct a witness to observe the violation of {\it{macroscopic no-signalling}}. 
\end{abstract}


\maketitle
\section{Introduction}

Bell's notion of local realism \cite{Bell, Bell66} aims for a classical understanding of the quantum world. As it turns out such a classical description is falsified in quantum theory. Macroscopic realism or macrorealism on the other hand aims to address the problem of whether macroscopic or simply classical systems can behave quantum mechanically or not. For instance, a famous problem in this regard is whether Schrodinger's cat can be alive and dead at the same time or not. Macrorealism embodies our classical understanding of the world, where macroscopic objects have well-defined states, and measurements of their properties do not cause any significant perturbations. These principles form the foundation of classical physics and have been deeply ingrained in our scientific and everyday thinking. It is well-known that in the realms of quantum physics, these classical intuitions are challenged, leading to the exploration of phenomena like quantum entanglement and non-locality. 

Macrorealism is based on two major assumptions. First, macrorealism per se: This principle suggests that a system, in the macroscopic sense, always exists in a well-defined and distinct state. In other words, for classical objects that we encounter in our everyday lives, there is an intrinsic property or state that fully characterizes the system at any given moment. This state is assumed to be definite, and it remains unchanged until acted upon by an external force or measurement. For instance consider a classical object, like a billiard ball on a pool table, macrorealism implies that it has a specific position and velocity, and this information uniquely defines its state.
Second, non-invasive measurements (NIM): This intuition posits that measurements made on a system do not interfere with or disturb the system. In classical physics, when one measures a property of an object, such as its position or velocity, one expects that the act of measurement doesn't alter these properties. The measurements are assumed to be passive observations, and the system remains in its original state after the measurement. Any system satisfying the above requirements, will be called macroscopic. For more details, refer to \cite{LGreview, extraref1, extraref2, extraref3, extraref4,extraref5, extraref6}. 

To observe the violation of macrorealism one needs to consider time-like correlations, that is, correlations between measurement events on a single system at different times. One can also comprehend the scenario like the standard Bell scenario but in the time domain. However, there are some crucial differences between both scenarios \cite{LGreview}. For instance, no-signalling is always satisfied in the Bell scenario but can be violated in the time domain. Consequently, no-signalling in time can serve as a witness to observe the violation of macrorealism \cite{Bruk2,Bruk3, Bruk4, Bruk5, Bruk1, Fritz_2010}. Even if no-signalling in time is satisfied one can still observe the violation of macrorealism using Leggett-Garg inequalities \cite{LG} which can be considered as a temporal version of Bell inequalities. It is important to note here that to observe a violation of Leggett-Garg inequalities, one requires at least three different measurements. Consequently, one can also obtain a violation of no-signaling in time even if Leggett-Garg inequalities are satisfied \cite{Bruk1}. It was further shown in \cite{mr33, mr34} that, unlike Bell inequalities in the spatial scenario, Leggett-Garg inequalities are not necessary to observe a violation of macrorealism. Thus, no-signalling in time is a better candidate to observe violation of macrorealism. In \cite{mr35}, it is further shown that LG inequalities primarily detect the violation of macrorealism per se and the NSIT tests detect the violation of NIM.

The violation of Leggett-Garg inequalities has been experimentally demonstrated but only for microscopic systems \cite{LGexp1, LGexp2, LGexp3,LGexp4,LGexp5, extrarefexp1, extrarfexp2, extrarefexp3, extrarefexp4, extrarefexp5}. It still remains a challenge to demonstrate the violation of Leggett-Garg inequalities in systems that can be considered macroscopic. The maximal violation of Leggett-Garg inequalities has also been utilised for semi-device independent certification of quantum measurements \cite{Jeba, Das, sarkar21}. Interestingly, it was shown in \cite{sarkar21} that one can utilise the maximal violation of Leggett-Garg inequalities for certification of an unbounded amount of randomness without nonlocality.

In this work, we show that even if no-signalling in time and Leggett-Garg inequalities are satisfied, still one can observe the violation of macrorealism. Consequently, none of the above mentioned conditions are necessary to observe a violation of macrorealism. For our purpose, we consider two space-like separated parties each of whom have access to shared randomness. Both the parties need to perform a single measurement. We then impose that the local correlations of the parties are no-signalling in time. We then observe that by using a single measurement with each party one can obtain a violation of macrorealism. Furthermore, we identify a larger class of classical models that can be falsified using the proposed setup. We term the corresponding notion as {\it{macroscopic no-signaling}}. We then construct a simple witness to observe the violation of macroscopic no-signaling. We then extend the setup to the case when the parties can perform measurements with arbitrary number of outcomes. This shows that violation of macroscopic no-signalling can be obtained by states and measurements acting on arbitrary dimensional Hilbert spaces. Furthermore, we identify some necessary conditions for the violation of macroscopic no-signalling.

\section{Scenario}

\begin{figure}[t!]
    \centering
    \includegraphics[scale=.45]{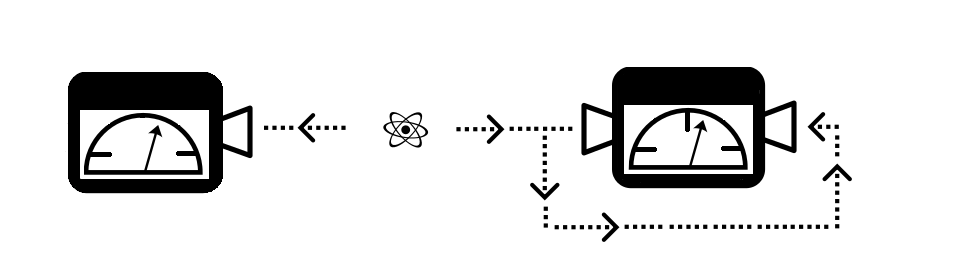}
    \caption{The setup involves two players, Bob and Alice, along with a preparation device that dispatches a subsystem to each of them. Bob and Alice are each limited to a single measurement on their respective subsystems. However, Alice is allowed to conduct same measurement twice on her subsystem. In this context, we confine Alice's measurements to two specific time points, denoted as $t_0$ and $t_1$. Bob and Alice obtain the joint probability distribution $\{p(a_0,a_1,b)\}$ at the end of the experiment.}
    \label{fig:enter-label}
\end{figure}

The scenario consists of two parties, namely, Bob and Alice, and a preparation device that sends one subsystem to Bob and one to Alice. Bob and Alice can only perform a single measurement on their subsystems. However, Alice can sequentially measure her subsystem, that is, she can twice apply the same measurement on her subsystem. Here, we restrict Alice to measure her system twice, that is, at times $t_0,t_1$. Now, Bob and Alice repeat the experiment enough times to construct the joint probability distribution (correlations) $\vec{p}=\{p(a_0,a_1,b)\}$ where $p(a_0,a_1,b)$ denotes the probability of obtaining outcome $a_0,a_1$ at times $t_0,t_1$ with Alice and $b$ with Bob respectively. Here we consider that $b=0,1$ and $a_0,a_1=0,1,2$. The scenario is depicted in Fig. \ref{fig:enter-label}. 

Let us now restrict to quantum theory and consider that the source sends the quantum state $\rho_{AB}$ with Bob and Alice performing the measurement $\{\mathcal{M}_i\}$ and $\{\mathcal{R}_i\}$ respectively, where $\mathcal{M}_i\geq0,\mathcal{R}_i\geq0$ and $\sum_i\mathcal{M}_i=\sum_i\mathcal{R}_i=\I$. As the measurements, in general, might not be projective, the probability $p(a_0,a_1,b)$ can be computed as \cite{niel}
\begin{equation}\label{probgen}
 p(a_0,a_1,b)=
\Tr\left(\sqrt{\mathcal{R}_{a_0}}U_{a_0}^{\dagger}\mathcal{R}_{a_1}U_{a_0}\sqrt{\mathcal{R}_{a_0}}\otimes\mathcal{M}_b\ \rho_{AB}\right)
\end{equation}
here $U_{a_0}$ is some unitary dependent on the outcome $a_0$. According to Luder's postulate \cite{Luder1, Luder2}, the unitaries in the above formula can be dropped.

We impose that Alice's local correlations obey the well-known "no-signalling in time (NSIT)" constraint \cite{LGreview}, which can be expressed as
\begin{eqnarray}\label{NSIT}
    \sum_{a_i}p(a_0,a_1)=p(a_j)\quad i\ne j.
\end{eqnarray}

\begin{figure}
    \centering
    \includegraphics[scale=.65]{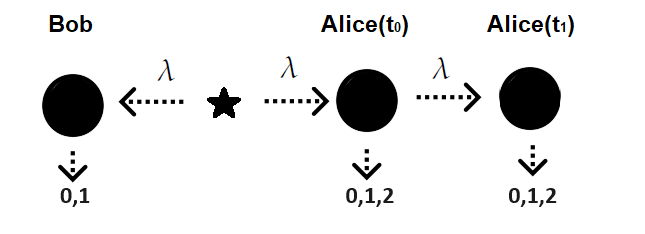}
    \caption{Macrorealistic description of the setup in Fig. \ref{fig:enter-label}. The source sends the variable $\lambda$ to both Bob and Alice with probability $p(\lambda)$. The variable does not change after Alice's measurement at time $t_0$.}
    \label{fig2}
\end{figure}

{\it{Macroscopic no-signalling}}--- Let us now specify the assumptions of {\it{macroscopic no-signalling}}. Consider again the setup depicted in Fig. \ref{fig:enter-label}. In general, one can consider that the source generates some variables $\lambda$ with probability $p(\lambda)$, which can treated as shared randomness between Bob and Alice [see Fig. \ref{fig2}]. In principle, $\lambda$ could also be some quantum states. Consequently, $p(a_0,a_1,b)$ can be expressed as
\begin{eqnarray}\label{eq1}
p(a_0,a_1,b)=\sum_{\lambda}p(a_0,a_1,b|\lambda)p(\lambda).
\end{eqnarray}
It is straightforward to see from Fig. \ref{fig2} that shared randomness can always be described locally, that is,
\begin{eqnarray}\label{eq2}
    p(a_0,a_1,b|\lambda)=p(a_0,a_1|\lambda)p(b|\lambda).
\end{eqnarray}
Now, macroscopic no-signalling can be simply stated as
\begin{defn}[Macroscopic no-signalling] Given any additional information $\lambda$, Alice's measurement outcome at time $t_0$ does not influence the outcome at a later time $t_1$, that is,
\begin{equation}\label{NSITO}
    \sum_{a_i}p(a_0,a_1|\lambda)=p(a_j|\lambda)\quad i\ne j \quad \forall \lambda.
\end{equation}
\end{defn}
Notice that the above definition encompasses a larger class of models than macroscopic realism which is defined as \cite{LG}
\begin{eqnarray}\label{mr}
     p(a_0,a_1|\lambda)=p(a_0|\lambda)p(a_1|\lambda).
\end{eqnarray}
As $\sum_{a_i} p(a_i|\lambda)=1$ for every $\lambda,i$, it is simple to observe that macroscopic realism Eq. \ref{mr} implies macroscopic no-signalling Eq. \ref{NSITO}, however, the converse is not true. Any correlation satisfying Eq. \eqref{mr} is termed as macroscopic no-signalling correlation.
Let us now observe the violation of macroscopic no-signalling.

\section{Violation of macroscopic no-signalling}

Consider again the correlations $p(a_0,a_1,b)$ written using Eq. \eqref{eq1} as
\begin{eqnarray}
\sum_{a_0}p(a_0,a_1,b)=\sum_{\lambda}\sum_{a_0}p(a_0,a_1,b|\lambda)p(\lambda).
\end{eqnarray}
Using Eq. \eqref{eq2} and then macroscopic no-signalling \eqref{NSITO} one can conclude that for all $a_1,b$
\begin{eqnarray}\label{MNSC}
\sum_{a_0}p(a_0,a_1,b)=\sum_{\lambda}p(a_1|\lambda)p(b|\lambda)p(\lambda)=p(a_1,b).
\end{eqnarray}
It should be noted here that although the above condition looks similar to the three-time NSIT conditions \cite{mr33}, both conditions are completely different. In the latter, one has to perform three sequential measurements on the same system and then observe a violation of NSIT conditions. However, here we impose that over the same system one always satisfies the NSIT conditions.

Consider now that the source prepares the classically correlated state 
\begin{eqnarray}\label{idealstate}
    \rho_{AB}=\frac{1}{2}\proj{0_A0_B}+\frac{1}{2}\proj{1_A1_B}
\end{eqnarray}
with Bob performing the measurement $\sigma_z$ and Alice performing the measurement
\begin{eqnarray}\label{idealmea}
  \mathcal{R}_0&=&\frac{1}{3}(\I+\sigma_z)\nonumber\\
   \mathcal{R}_1&=&\frac{1}{3}\left(\I-\frac{1}{2}\sigma_z+\frac{\sqrt{3}}{2}\sigma_x\right)\nonumber\\
     \mathcal{R}_2&=&\frac{1}{3}\left(\I-\frac{1}{2}\sigma_z-\frac{\sqrt{3}}{2}\sigma_x\right)
\end{eqnarray}
where $\sigma_z,\sigma_x$ are the Pauli $z,x$ observables respectively. For a note, the above measurement is also termed as {\it{trine-POVM}} and has been experimentally implemented \cite{POVM1}.
As the above Alice's measurement is an extremal POVM \cite{peri}, that is, each measurement element can be expressed as a projector times a positive number, the formula Eq. \eqref{probgen} is simplified to
\begin{eqnarray}
    p(a_0,a_1,b)=
\Tr\left(\sqrt{\mathcal{R}_{a_0}}\mathcal{R}_{a_1}\sqrt{\mathcal{R}_{a_0}}\otimes\mathcal{M}_b\ \rho_{AB}\right).
\end{eqnarray}
Notice that the local state of Alice is the maximally mixed state and thus for any measurement by Alice, the no-signalling in time condition \eqref{NSIT} will be satisfied. Substituting Bob's and Alice's measurements in the above formula allows us to observe that the condition \eqref{MNSC} is not satisfied for any $a_1,b$. Consequently, we can conclude that quantum theory violates the notion of macroscopic no-signalling with a single measurement with Alice.

{\it{Witness---}} From an experimental perspective, one can never observe a strict equality condition as Eq. \eqref{MNSC}. Thus, we now find a witness that allows one to observe a considerable difference between macroscopic no-signalling correlations and quantum correlations. The witness is given by

\begin{eqnarray}\label{Wit1}
\mathcal{S}=\sum_{a_1=0,1,2}\sum_{b=0,1}\left|\sum_{a_0=0,1,2}p(a_0,a_1,b)-p(a_1,b)\right|.
\end{eqnarray}
From Eq. \eqref{MNSC}, it is clear that for macroscopic no-signalling correlations $\mathcal{S}=0$. Using the measurements suggested in Eq. \eqref{idealmea} and the classically correlated state Eq. \eqref{idealstate}, one can attain a value $\mathcal{S}=1/3$. 

\section{Extension to arbitrary outcome measurements and high dimensional states}

Let us now extend the setup in Fig. \ref{fig:enter-label} to the case when Alice chooses a $2d-$outcome measurement with Bob performing a $d-$outcome one. Consequently, we have that $a_0,a_0=0,1,\ldots,2d-1$ and $b=0,1,\ldots,d-1$ with $d$ being an arbitrary positive integer such that $d\geq2$. Now, similar to the witness \eqref{Wit1}, we construct another witness to observe macroscopic no-signalling for arbitrary outcome measurements as
\begin{eqnarray}\label{Witd}
\mathcal{S}_d=\sum_{a_1=0}^{2d-1}\sum_{b=0}^{d-1}\left|\sum_{a_0=0}^{2d-1}p(a_0,a_1,b)-p(a_1,b)\right|.
\end{eqnarray}
Again, from Eq. \eqref{MNSC}, it is clear that for macroscopic no-signalling correlations $\mathcal{S}_d=0$. 

Consider now the following measurements with Alice $\mathcal{R}=\{\mathcal{R}_i\}$ where
\begin{eqnarray}\label{Alicemead}
    \mathcal{R}_i=
    \begin{cases}
        \frac{1}{2}\proj{i}\qquad i=0,1,\ldots,d-1\nonumber\\
 \frac{1}{2}\proj{\mu_i}\quad i=d,\ldots,2d-1
    \end{cases}
\end{eqnarray}
such that $\{\ket{i}\}$ is the computational basis and $\{\ket{\mu_i}\}$ denote the Hadamard basis, that is, 
\begin{eqnarray}
    \ket{\mu_i}=\frac{1}{\sqrt{d}}\sum_{k=0}^{d-1}\omega^{(2d-i)k}\ket{k}.
\end{eqnarray}
Let us also notice that $\sum_{i=0}^{d-1}\mathcal{R}_i=\sum_{i=d}^{2d-1}\mathcal{R}_i=\I/2$ along with the fact that $|\bra{\mu_i}j\rangle|^2=1/d$ for any $i,j$. Moreover, let us also consider that Bob again performs his measurement in the computational basis and the source generates the classically correlated state $\rho^D_{AB}$
\begin{eqnarray}\label{idealstated}
    \rho_{AB}^d=\frac{1}{2}\sum_{i=0}^{d-1}\proj{i_Ai_B}.
\end{eqnarray}

Notice again that the local state of Alice is the maximally mixed state and thus for any measurement by Alice, the no-signalling in time condition \eqref{NSIT} will be satisfied. Let us now compute the value of the witness $\mathcal{S}_d$ \eqref{Witd} using the above-mentioned state and measurements. For this purpose, let us first observe that Alice's measurement \eqref{Alicemead} for any $a_1$ satisfies 
\begin{eqnarray}
    \sum_{a_0=0}^{2d-1}\sqrt{\mathcal{R}_{a_0}}\mathcal{R}_{a_1}\sqrt{\mathcal{R}_{a_0}}=\frac{\I}{4d}+\mathcal{R}_{a_1}^2.
\end{eqnarray}
This allows us to obtain from \eqref{Witd}
\begin{eqnarray}
    \mathcal{S}_d=\frac{1}{2}\sum_{a_1=0}^{2d-1}\sum_{b=0}^{d-1}\left|\frac{1}{2d^2}-p(a_1,b)\right|.
\end{eqnarray}
Evaluating $p(a_1,b)$ for all $a_1,b$ gives us 
\begin{eqnarray}
    \mathcal{S}_d=\frac{1}{2}\left(1-\frac{1}{d}\right).
\end{eqnarray}

\section{Necessary conditions for violation}
Let us now identify some necessary conditions to violate macroscopic no-signalling. For this purpose, let us consider the correlation $p(a_0,a_1,b)$ as in Eq. \eqref{probgen}. Assuming Luder's postulate, let us find the properties of Alice's measurement $\{\mathcal{R}_i\}$ that can satisfy the condition \eqref{MNSC}. Now, expanding \eqref{MNSC} gives us
\begin{equation}\label{mr1}
\sum_{a_0}\Tr\left(\sqrt{\mathcal{R}_{a_0}}\mathcal{R}_{a_1}\sqrt{\mathcal{R}_{a_0}}\otimes\mathcal{M}_b\ \rho_{AB}\right)=\Tr\left(\mathcal{R}_{a_1}\otimes\mathcal{M}_b\ \rho_{AB}\right).
\end{equation}
It is clear from the above expression that if $\mathcal{R}_{a_0}$ and $\mathcal{R}_{a_1}$ for any $a_0,a_1$ commute, then the above relation \eqref{mr1} is always satisfied. Thus, a necessary condition to observe a violation of macroscopic no-signalling is that Alice's measurement elements must be incompatible with each other. This rules out two major classes of measurements,
\begin{enumerate}
    \item As projectors always commute among each other, macroscopic no-signalling can not be violated using projective measurements with Alice.
    \item The measurement elements of any two-outcome measurement also commute among each other. Consequently, to observe a violation of macroscopic no-signalling, Alice needs to perform a measurement with at least three outcomes.
\end{enumerate}
Consequently, we can also define a class of measurements where the measurement elements commute among each other.

\begin{defn}[Self-commuting measurements] Consider a measurement $\mathcal{R}=\{\mathcal{R}_i\}$ such that every measurement element commutes among each other, that is, $\mathcal{R}_i\mathcal{R}_j=\mathcal{R}_j\mathcal{R}_i$ for any $i,j$. Then the measurement $\mathcal{R}$ is a self-commuting measurement.   
\end{defn}

Self-commuting measurements can not be used to violate the notion of macroscopic no-signalling. It will be interesting to find further properties of non-projective measurements that can violate macroscopic no-signalling.

\section{Conclusions}

The above-presented result can also be considered as a violation of Leibniz's principle of indiscernibles \cite{sep}, which can be simply stated as principles that hold operationally should also hold ontologically. Here, no-signalling in time is satisfied operationally, but is violated at the ontological level. One might argue that the notion of macrorealism or macroscopic no-signalling is very strong even for macroscopic systems. It is important to note that in the presence of memory, even classical systems can violate these notions \cite{sarkar21}. Consequently, in the above work, we are comparing non-classical versus classical theories in absence of memory.  For a note, a weaker form of macrorealism imposed on the ontological description of quantum systems, termed ontic-distinguishability was considered in \cite{Sarkarfoop}. 

Several follow-up problems arise from this work. The most interesting among them would be to find a unifying notion of local realism and macrorealism using a similar setup to the one presented in this work. 
Finding the maximal value of $\mathcal{S}$ \eqref{Wit1} or $\mathcal{S}_d$ \eqref{Witd} is a challenging task given the fact that both these functionals are non-linear, and one would have to optimise over all possible states and measurements. The construction of a linear inequality to observe the violation of macroscopic no-signalling is thus highly desirable. As the witness suggested in this work is simple and can be violated using a classically correlated state and the trine-POVM \eqref{idealmea} which has been experimentally implemented \cite{POVM1}, we believe that it will not be difficult to observe the violation of macroscopic no-signalling at least for microscopic quantum systems.

\begin{acknowledgments}
This project was funded within the QuantERA II Programme (VERIqTAS project) that has received funding from the European Union’s Horizon 2020 research and innovation programme under Grant Agreement No 101017733.
\end{acknowledgments}

\providecommand{\noopsort}[1]{}\providecommand{\singleletter}[1]{#1}%

\begin{figure*}
    \centering
    \includegraphics[width=\linewidth]{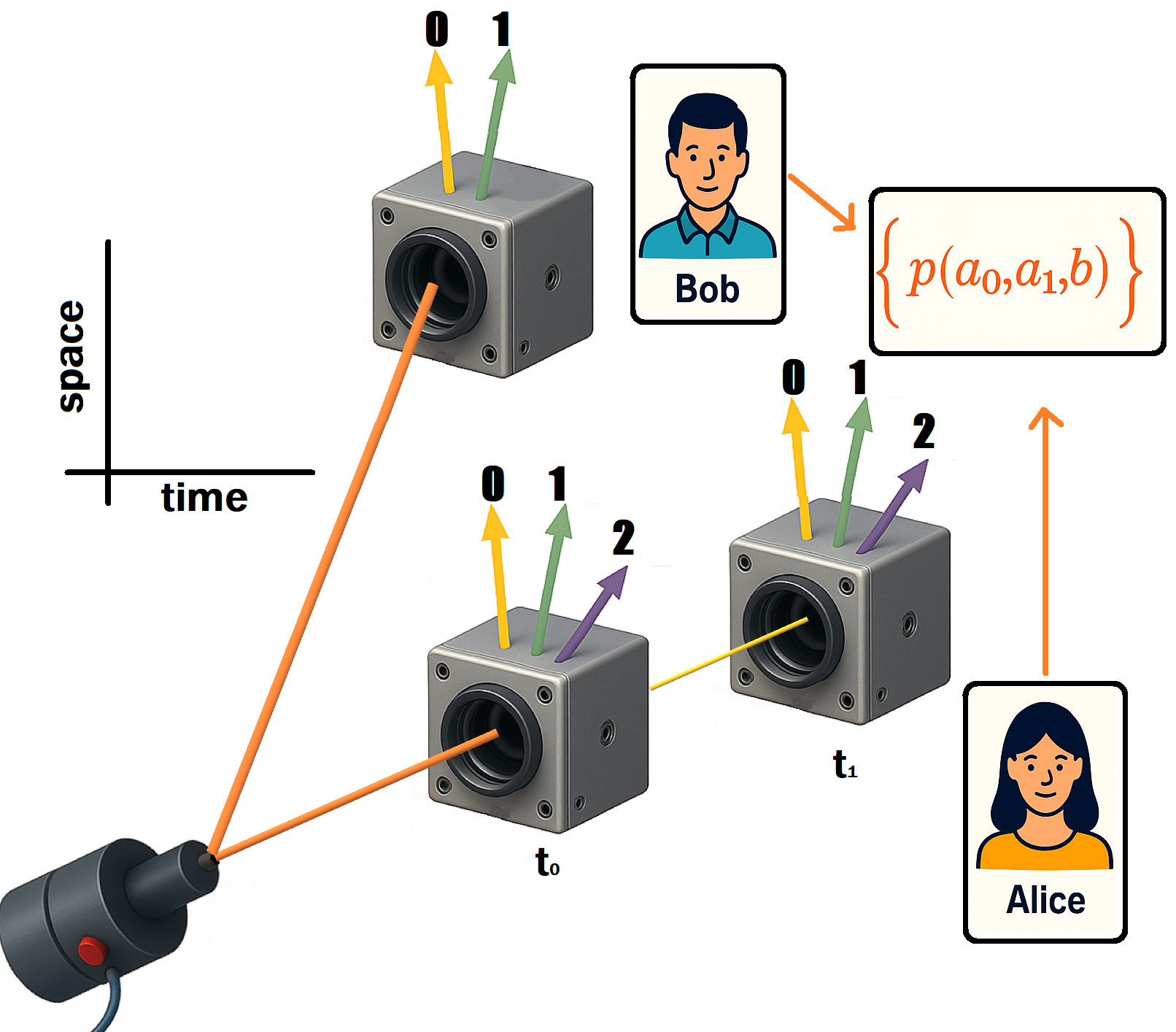}
    \caption{A beautified image of Fig.1}
    \label{11}
\end{figure*}

\end{document}